\documentclass[fleqn, 11pt]{article}
\usepackage{amsmath, amssymb, amsthm, graphicx}
\theoremstyle{definition}
\newtheorem{theorem}{Theorem}[section]

\newtheorem{Remark}[theorem]{Remark}

\newtheorem{Example}[theorem]{Example}
\newtheorem{Question}[theorem]{Question}

\numberwithin{equation}{section}

\newcommand{\captionfonts}{\small}
\makeatletter 
\long\def\@makecaption#1#2{%
\vskip\abovecaptionskip
\sbox\@tempboxa{{\captionfonts #1: #2}}%
\ifdim \wd\@tempboxa >\hsize {\captionfonts #1: #2\par} \else
\hbox to\hsize{\hfil\box\@tempboxa\hfil}%
\fi \vskip\belowcaptionskip}
\makeatother 

\title{Finiteness of polygonal relative equilibria for generalised quasi-homogeneous $n$-body problems and $n$-body problems in spaces of constant curvature}
\author{Pieter Tibboel}
\begin{document}
\maketitle
\begin{abstract}
  We prove for generalisations of quasi-homogeneous $n$-body problems with center of mass zero and $n$-body problems in spaces of negative constant Gaussian curvature that if the masses and rotation are fixed, there exists, for every order of the masses, at most one equivalence class of relative equilibria for which the point masses lie on a circle, as well as that there exists, for every order of the masses, at most one equivalence class of relative equilibria for which all but one of the point masses lie on a circle and rotate around the remaining point mass. The method of proof is a generalised version of a proof by J.M. Cors, G.R. Hall and G.E. Roberts on the uniqueness of co-circular central configurations for power-law potentials.
\end{abstract}
  \section{Introduction}
  By $n$-body problems we mean problems where we are tasked with deducing the dynamics of $n$ point masses described by a system of differential equations. The study of such problems has applications to various fields, including atomic physics, celestial mechanics, chemistry, crystallography, differential equations, dynamical systems, geometric mechanics, Lie groups and algebras, non-Euclidean and differential geometry, stability theory, the theory of polytopes and topology (see for example \cite{AbrahamMarsden}, \cite{AlbouyCabralSantos}, \cite{D2}, \cite{D4}, \cite{D6}, \cite{DPS4}, \cite{Saari}, \cite{Smale}, \cite{Smale 2}, \cite{Smale2}, \cite{Wintner} and the references therein). The $n$-body problems that form the backbone of this paper are a generalisation of a class of quasi-homogeneous $n$-body problems, which we will call generalised $n$-body problems for short and the $n$-body problem in spaces of constant Gaussian curvature, or curved $n$-body problem for short. By the generalised $n$-body problem we mean the problem of finding the orbits of  point masses $q_{1}$,...,$q_{n}\in\mathbb{R}^{2}$ and respective masses $m_{1}>0$,...,$m_{n}>0$ determined by the system of differential equations
  \begin{align}\label{Equations of motion}
    \ddot{q}_{i}=\sum\limits_{j=1,\textrm{ }j\neq i}^{n}m_{j}(q_{j}-q_{i})f\left(\|q_{j}-q_{i}\|\right),
  \end{align}
  where $\|\cdot\|$ is the Euclidean norm, $f$ is a positive valued scalar function and $xf(x)$ is a decreasing, differentiable function. Our definition of generalised $n$-body problems thus includes a large subset of quasi-homogeneous $n$-body problems, which are problems with $f(x)=Ax^{-a}+Bx^{-b}$, where $A$, $B\in\mathbb{R}$ and $0\leq a<b$, which include problems studied in fields such as celestial mechanics, crystallography, chemistry and electromagnetics (see for example \cite{CLP}--\cite{D0}, \cite{DDLMMPS}, \cite{DMS}, \cite{DPS0},  \cite{DPS4}, \cite{J} and \cite{P}--\cite{PV}).

  By the $n$-body problem in spaces of constant Gaussian curvature, we mean the problem of finding the dynamics of point masses \begin{align*}p_{1},...,p_{n}\in\mathbb{M}_{\sigma}^{2}=\{(x_{1},x_{2},x_{3})\in\mathbb{R}^{3}|x_{1}^{2}+x_{2}^{2}+\sigma x_{3}^{2}=\sigma\},\end{align*} where $\sigma=\pm 1$ and respective masses $\widehat{m}_{1}>0$,...,$\widehat{m}_{n}>0$, determined by the system of differential equations
  \begin{align}\label{EquationsOfMotion Curved}
   \ddot{p}_{i}=\sum\limits_{j=1,\textrm{ }j\neq i}^{n}\frac{\widehat{m}_{j}(p_{j}-\sigma(p_{i}\odot p_{j})p_{i})}{(\sigma -\sigma(p_{i}\odot p_{j})^{2})^{\frac{3}{2}}}-\sigma(\dot{p}_{i}\odot\dot{p}_{i})p_{i},\textrm{ }i\in\{1,...,n\}.
  \end{align}
  where for $x$, $y\in\mathbb{M}_{\sigma}^{2}$  the product $\cdot\odot\cdot$ is defined as
  \begin{align*}
    x\odot y=x_{1}y_{1}+x_{2}y_{2}+\sigma x_{3}y_{3}.
  \end{align*}
  The curved $n$-body problem generalises the classical, or Newtonian $n$-body problem ($f(x)=x^{-\frac{3}{2}}$ in (\ref{Equations of motion})) to spaces of constant Gaussian curvature (i.e. spheres and hyperboloids) and goes for the two body case back to the 1830s, (see \cite{BB} and \cite{Lo}), followed by \cite{S1}, \cite{S2}, \cite{K1}, \cite{K2}, \cite{K3}, \cite{L1}, \cite{L2}, \cite{L3}, \cite{KH}, but it was not until a revolution took place with the papers \cite{DPS1}, \cite{DPS2}, \cite{DPS3}  by Diacu, P\'erez-Chavela and Santoprete in which the succesful study of $n$-body problems in spaces of constant Gaussian curvature for the case that $n\geq 2$ was established. After this breakthrough, further results for the $n\geq 2$ case were then obtained in \cite{CRS}, \cite{D1}--\cite{D5}, \cite{DK}, \cite{DP}, \cite{DPo}, \cite{DT} and \cite{T}--\cite{T4}. See \cite{D1}--\cite{D4} and \cite{DK} for a detailed historical overview of the development of the curved $n$-body problem. In this paper we will only consider the negative constant curvature case, i.e. the case $\sigma=-1$.

  For these two types of $n$-body problems we will prove results regarding the finiteness of relative equilibrium solutions of (\ref{Equations of motion}) and (\ref{EquationsOfMotion Curved}), which are solutions of (\ref{Equations of motion}), or (\ref{EquationsOfMotion Curved}), for which the configuration of the point masses stays fixed in shape and size over time. Specifically:

  We will call $q_{1}$,...,$q_{n}\in\mathbb{R}^{2}$ a \textit{relative equilibrium} of (\ref{Equations of motion}) if
 $q_{i}(t)=T(At)(Q_{i}-Q_{M})+Q_{M}$, $i\in\{1,...,n\}$,
  where $Q_{i}\in\mathbb{R}^{2}$, $A\in\mathbb{R}_{>0}$ are constant,
  \begin{align*}
    T(t)=\begin{pmatrix}
      \cos{t} & -\sin{t} \\
      \sin{t} & \cos{t}
    \end{pmatrix}
  \end{align*}
    is a $2\times 2$ rotation matrix and
    \begin{align*}
      Q_{M}=\frac{1}{M}\sum\limits_{k=1}^{n}m_{k}Q_{k}
    \end{align*}
    is the \textit{center of mass} with $M=\sum\limits_{k=1}^{n}m_{k}$. If the $q_{i}$ lie on a circle with the origin at its center, we will call $q_{1}$,...,$q_{n}$ a polygonal relative equilibrium solution of (\ref{Equations of motion}). If all but one of the masses form a polygon with the origin at its center, with the remaining mass at the origin, then we will call such a relative equilibrium a polygonal relative equilibrium with center zero of (\ref{Equations of motion}) for short.

  Following the example of \cite{DPS1}, \cite{DPS2}, \cite{DPS3}  by Diacu, P\'erez-Chavela and Santoprete, we will call $p_{1}$,...,$p_{n}\in\mathbb{M}_{\sigma}^{2}$ a polygonal relative equilibrium of (\ref{EquationsOfMotion Curved}) if
  \begin{align*}
    p_{i}(t)=\begin{pmatrix}
      T(Bt)P_{i} \\
      z
    \end{pmatrix},
  \end{align*}
  where $P_{i}\in\mathbb{R}^{2}$, $z\in\mathbb{R}$, $B\in\mathbb{R}_{>0}$ are constant and $i\in\{1,...,n\}$. For a proof of the existence of such solutions we refer the reader to Theorem 1 of \cite{D2}.

  Finally, following \cite{DPS4}, we will say that two relative equilibria of (\ref{Equations of motion}) are equivalent, or are in the same equivalence class, if they are equivalent under rotation. For the constant curvature case, we will say that two polygonal relative equilibria are equivalent if they are equivalent under a rotation induced by a rotation matrix of the type $\begin{pmatrix}
    T(c) & \mathbf{0} \\
    \mathbf{0}^{T} & 1
  \end{pmatrix}$, where $c\in\mathbb{R}$ is a constant, $\mathbf{0}\in\mathbb{R}^{2}$ is the zero vector and $\mathbf{0}^{T}$ its transpose. It should be noted that these definitions differ from the usual definition (see for example \cite{Smale}), where two relative equilibria are also considered to be equivalent if they are equivalent under scalar multiplication.

  The relevance of the relative equilibria studied in this paper is twofold: In \cite{CHG}, Cors, Hall and Roberts proved for the case that if $Q_{M}=0$, $f(x)=x^{-\alpha-2}$, $\alpha>0$ and if $A$ and the masses $m_{1}$,...,$m_{n}$ are fixed, then for every order of the masses there exists at most one equivalence class of polygonal relative equilibrium solutions (called co-circular central configurations for power-law potentials in \cite{CHG}) of (\ref{Equations of motion}). This may be a significant step in the direction of proving Problem 12 of \cite{AlbouyCabralSantos}, an important list of open problems in the field of celestial mechanics, composed by Albouy, Cabral and Santos:  Are there, except for the regular $n$-gon with equal masses, any polygonal relative equilibria of (\ref{Equations of motion}) for the case that $f(x)=x^{-a}$, $a\geq 1$? A logical step to make is to investigate to which extent Cors', Hall's and Roberts' result can be applied to $n$-body problems in spaces of constant curvature, or any generalised $n$-body problems. Additionally, generalising this result may shed further light on solving Problem 12 of \cite{AlbouyCabralSantos} and the sixth Smale problem, which conjectures that for any fixed set of masses, the corresponding set of equivalence classes of relative equilibria of the classical $n$-body problem is finite (see \cite{Smale}). Secondly and entwined with the theoretical aspect, relative equilibria can tell us a great deal about the geometry of the universe and orbits in our solar system: It was proven in \cite{DPS1} and \cite{DPS2} that while for the zero curvature case polygonal relative equilibria shaped as equilateral triangles with unequal masses exist, in nonzero constant curvature spaces the masses have to be equal, proving that the region between the Sun, Jupiter and the Trojan asteroids has to be flat. This means that getting any information about polygonal relative equilibria that exist in spaces of positive constant curvature, zero curvature, or negative constant curvature can further our understanding about the geometry of the universe. Additionally, the ring problem, or a regular polygonal relative equilibrium with one mass at its center and all masses on the circle equal (see for example \cite{GL}) is a model that was originally formulated by Maxwell to describe the dynamics of particles orbiting Saturn (see \cite{M}) and has since then been applied to describing other planetary rings, asteroid belts, planets orbiting stars, stellar formations, stars with an accretion ring, planetary nebula and motion of satellites (see \cite{AE}, \cite{AEKP}, \cite{AEP}, \cite{GL}, \cite{HK}, \cite{Ka}, \cite{Ka2}, \cite{MS}, \cite{MS2},  \cite{Moeckel}, \cite{Saari}--\cite{Sch2}). In this context, considering the more general solutions of polygonal relative equilibria, proving the number of possible equilibria to be finite may be a very fruitful endeavour. We will prove the following theorems:
   \begin{theorem}\label{Main Theorem}
    Let $A$, $m_{1}$,...,$m_{n}$ be fixed. For every order of the masses, there exists at most one equivalence class of polygonal relative equilibria of (\ref{Equations of motion}) with $Q_{M}=0$.
   \end{theorem}
   \begin{theorem}\label{Main Theorem 2}
     Let $B$, $\widehat{m}_{1}$,...,$\widehat{m}_{n}$ be fixed and let $\sigma=-1$. For every order of the masses, there exists at most one equivalence class of polygonal relative equilibria of (\ref{EquationsOfMotion Curved}). 
   \end{theorem}
   \begin{theorem}\label{Main Theorem 3}
     Let $A$, $m_{1}$,...,$m_{n}$ be fixed. Let $n=N+1$. Then for every order of the masses, there exists at most one equivalence class of relative equilibria of (\ref{Equations of motion}) with $Q_{M}=0$ where for $i\in\{1,...,N\}$ the $q_{i}$ lie on a circle with the origin at its center and $q_{N+1}=0$.
   \end{theorem}
  We will first prove Theorem~\ref{Main Theorem} in section~\ref{Proof of main theorem}, after which we will prove Theorem~\ref{Main Theorem 2} in section~\ref{Proof of main theorem 2} and finally Theorem~\ref{Main Theorem 3} in section~\ref{Proof of main theorem 3}.
  \section{Proof of Theorem~\ref{Main Theorem}}\label{Proof of main theorem}
  We will prove Theorem~\ref{Main Theorem} by to a large extent following the proof of Theorem~3.2 in \cite{CHG}, which is reminiscent of a topological approach by Moulton (see \cite{Moulton}, \cite{CHG}), but instead of making the proof work for $f(x)=x^{-\alpha-2}$, where $\alpha>0$, as was done in \cite{CHG}, we successfully realise the result for any positive function $f$ for which $xf(x)$ is a decreasing function:

  If $q_{i}$, $i\in\{1,...,n\}$ is a relative equilibrium of (\ref{Equations of motion}) and $Q_{M}=0$, we may write $q_{i}(t)=T(At)Q_{i}$, where
  \begin{align*}
    Q_{i}=r\begin{pmatrix}
      \cos{\alpha_{i}}\\
      \sin{\alpha_{i}}
    \end{pmatrix},\textrm{ for } i\in\{1,...,n\},\textrm{ }r>0
  \end{align*}
  and $0\leq\alpha_{1}<\alpha_{2}<...<\alpha_{n}<2\pi$, meaning that if we insert $q_{i}(t)=rT(At)Q_{i}$ and $q_{j}(t)=rT(At)Q_{j}$ into (\ref{Equations of motion}) and using that in that case \begin{align*}\|Q_{i}-Q_{j}\|=r\sqrt{2-2\cos{(\alpha_{i}-\alpha_{j})}}=2r\sin{\left(\frac{1}{2}|\alpha_{i}-\alpha_{j}|\right)}\end{align*} and multiplying both sides of the resulting equation with $-T(-\alpha_{i})$ from the left, we can rewrite (\ref{Equations of motion}) as
  \begin{align}\label{Identity relative equilibria}
    r\begin{pmatrix}
      A^{2}\\ 0
    \end{pmatrix}=r\sum\limits_{j=1,\textrm{ }j\neq i}^{n}m_{j}\begin{pmatrix}
      1-\cos{(\alpha_{i}-\alpha_{j})}\\ \sin{(\alpha_{i}-\alpha_{j})}
    \end{pmatrix}f\left(2r\sin{\left(\frac{1}{2}|\alpha_{i}-\alpha_{j}|\right)}\right),
  \end{align}
  which, if we write $g(x)=xf(x)$, can be rewritten as
  \begin{align}\label{Identity relative equilibria3}
    \begin{cases}
      m_{i}A^{2}r=\sum\limits_{j=1,\textrm{ }j\neq i}^{n}m_{i}m_{j}
      \sin{\left(\frac{1}{2}|\alpha_{i}-\alpha_{j}|\right)}
    g\left(2r\sin{\left(\frac{1}{2}|\alpha_{i}-\alpha_{j}|\right)}\right)\\
    0=\sum\limits_{j=1,\textrm{ }j\neq i}^{n}m_{i}m_{j}r\delta_{ij}\cos{\left(\frac{1}{2}(\alpha_{i}-\alpha_{j})\right)}g\left(2r\sin{\left(\frac{1}{2}|\alpha_{i}-\alpha_{j}|\right)}\right),
    \end{cases}
  \end{align}
  where
  \begin{align*}
    \delta_{ij}=\begin{cases}
      1\textrm{ if }i>j \\
      -1\textrm{ if }i<j.
    \end{cases}
  \end{align*}
  If $G$ is any scalar function for which $G'(x)=g(x)$, then defining
  \begin{align}\label{definition V}
    V(r,\alpha_{1},...,\alpha_{n})=\sum\limits_{l=1}^{n}\sum\limits_{k=1,\textrm{ }k\neq l}^{n}m_{l}m_{k}G\left(2r\sin{\left(\frac{1}{2}|\alpha_{l}-\alpha_{k}|\right)}\right)-\sum\limits_{l=1}^{n}m_{l}A^{2}r^{2}
  \end{align}
  gives by (\ref{Identity relative equilibria3}) that
  \begin{align*}
    \frac{\partial V}{\partial r}=0\textrm{ and }\frac{\partial V}{\partial\alpha_{i}}=0\textrm{ for all }i\in\{1,...,n\}.
  \end{align*}
  This means that, for whatever values of $r$, $\alpha_{1}$,...,$\alpha_{n}$ the vectors $q_{i}(t)=T(At)Q_{i}$ give a relative equilibrium solution of (\ref{Equations of motion}), $(r,\alpha_{1},...,\alpha_{n})$ is a stationary point of $V$. We will show that for such a stationary point $V$ has to have a local maximum, i.e.
  \begin{align}\label{THE INEQUALITY}
    \rho^{2}\frac{\partial^{2}V}{\partial r^{2}}+2\rho\sum\limits_{l=1}^{n}\gamma_{l}\frac{\partial^{2}V}{\partial r\partial\alpha_{l}}+\sum\limits_{l=1}^{n}\sum_{k=1}^{n}\gamma_{l}\gamma_{k}\frac{\partial^{2}V}{\partial\alpha_{l}\partial\alpha_{k}}\leq 0
  \end{align}
  for all vectors $(\rho,\gamma_{1},...,\gamma_{n})\in\mathbb{R}^{n+1}$. \\
  Note that by (\ref{Identity relative equilibria3})
  \begin{align}\label{First part of THE INEQUALITY}
    \rho^{2}\frac{\partial^{2}V}{\partial r^{2}}&=4\rho^{2}\sum\limits_{i=1}^{n}\sum\limits_{j=1,\textrm{ }j\neq i}^{n}m_{i}m_{j}
      \sin^{2}{\left(\frac{1}{2}|\alpha_{i}-\alpha_{j}|\right)}
    g'\left(2r\sin{\left(\frac{1}{2}|\alpha_{i}-\alpha_{j}|\right)}\right)-2\rho^{2}\sum\limits_{i=1}^{n}m_{i}A^{2}.
  \end{align}
  Secondly, again by (\ref{Identity relative equilibria3}), note that
  \begin{align*}
    2\rho\sum\limits_{i=1}^{n}\gamma_{i}\frac{\partial^{2}V}{\partial r\partial\alpha_{i}}=4\rho\sum\limits_{i=1}^{n}\gamma_{i}\frac{\partial}{\partial r}\sum\limits_{j=1,\textrm{ }j\neq i}^{n}m_{i}m_{j}r\delta_{ij}\cos{\left(\frac{1}{2}(\alpha_{i}-\alpha_{j})\right)}g\left(2r\sin{\left(\frac{1}{2}|\alpha_{i}-\alpha_{j}|\right)}\right) \\
    =4\rho\sum\limits_{i=1}^{n}\gamma_{i}\left(\sum\limits_{j=1,\textrm{ }j\neq i}^{n}m_{i}m_{j}\delta_{ij}\cos{\left(\frac{1}{2}(\alpha_{i}-\alpha_{j})\right)}g\left(2r\sin{\left(\frac{1}{2}|\alpha_{i}-\alpha_{j}|\right)}\right)\right. \\
    +\left.2\sum\limits_{j=1,\textrm{ }j\neq i}^{n}m_{i}m_{j}r\cos{\left(\frac{1}{2}(\alpha_{i}-\alpha_{j})\right)}\sin{\left(\frac{1}{2}(\alpha_{i}-\alpha_{j})\right)}g'\left(2r\sin{\left(\frac{1}{2}|\alpha_{i}-\alpha_{j}|\right)}\right) \right),
  \end{align*}
  which, by the second identity of (\ref{Identity relative equilibria3}), gives
  \begin{align*}
    &2\rho\sum\limits_{i=1}^{n}\gamma_{i}\frac{\partial^{2}V}{\partial r\partial\alpha_{i}}=\nonumber\\
    &4\rho\sum\limits_{i=1}^{n}\left(0+2\gamma_{i}\sum\limits_{j=1,\textrm{ }j\neq i}^{n}m_{i}m_{j}r\cos{\left(\frac{1}{2}(\alpha_{i}-\alpha_{j})\right)}\sin{\left(\frac{1}{2}(\alpha_{i}-\alpha_{j})\right)}g'\left(2r\sin{\left(\frac{1}{2}|\alpha_{i}-\alpha_{j}|\right)}\right) \right)\\
    &=4\rho\sum\limits_{i=1}^{n}\sum\limits_{j=1,\textrm{ }j\neq i}^{n}(\gamma_{i}-\gamma_{j})m_{i}m_{j}r\cos{\left(\frac{1}{2}(\alpha_{i}-\alpha_{j})\right)}\sin{\left(\frac{1}{2}(\alpha_{i}-\alpha_{j})\right)}g'\left(2r\sin{\left(\frac{1}{2}|\alpha_{i}-\alpha_{j}|\right)}\right),
  \end{align*}
  so combined with (\ref{First part of THE INEQUALITY}), this gives
  \begin{align*}
    &\rho^{2}\frac{\partial^{2}V}{\partial r^{2}}+2\rho\sum\limits_{i=1}^{n}\gamma_{i}\frac{\partial^{2}V}{\partial r\partial\alpha_{i}}=-2\rho^{2}\sum\limits_{i=1}^{n}m_{i}A^{2}+\sum\limits_{i=1}^{n}\sum\limits_{j=1,\textrm{ }j\neq i}^{n}m_{i}m_{j}\left(4\rho^{2}\sin^{2}{\left(\frac{1}{2}|\alpha_{i}-\alpha_{j}|\right)}\right.\\
    &\left.+4(\gamma_{i}-\gamma_{j})\rho r\cos{\left(\frac{1}{2}(\alpha_{i}-\alpha_{j})\right)}\sin{\left(\frac{1}{2}(\alpha_{i}-\alpha_{j})\right)}\right)g'\left(2r\sin{\left(\frac{1}{2}|\alpha_{i}-\alpha_{j}|\right)}\right)
  \end{align*}
  which can be rewritten as
  \begin{align}\label{First and second part of THE INEQUALITY}
    &\rho^{2}\frac{\partial^{2}V}{\partial r^{2}}+2\rho\sum\limits_{i=1}^{n}\gamma_{i}\frac{\partial^{2}V}{\partial r\partial\alpha_{i}}=-2\rho^{2}\sum\limits_{i=1}^{n}m_{i}A^{2}\nonumber\\
    &+\sum\limits_{i=1}^{n}\sum\limits_{j=1,\textrm{ }j\neq i}^{n}m_{i}m_{j}\left(2\rho\sin{\left(\frac{1}{2}(\alpha_{i}-\alpha_{j})\right)}+r(\gamma_{i}-\gamma_{j})\cos{\left(\frac{1}{2}(\alpha_{i}-\alpha_{j})\right)}\right)^{2}\nonumber\\
    &\cdot g'\left(2r\sin{\left(\frac{1}{2}|\alpha_{i}-\alpha_{j}|\right)}\right)\nonumber\\
    &-\sum\limits_{i=1}^{n}\sum\limits_{j=1,\textrm{ }j\neq i}^{n}m_{i}m_{j}(\gamma_{i}-\gamma_{j})^{2}r^{2}\cos^{2}{\left(\frac{1}{2}(\alpha_{i}-\alpha_{j})\right)}g'\left(2r\sin{\left(\frac{1}{2}|\alpha_{i}-\alpha_{j}|\right)}\right)
  \end{align}
  Thirdly, for $i\neq j$, by (\ref{Identity relative equilibria3}),
  \begin{align*}
    \frac{\partial^{2}V}{\partial\alpha_{i}\partial\alpha_{j}}&=2\frac{\partial}{\partial\alpha_{j}}\left(m_{i}m_{j}r\delta_{ij}\cos{\left(\frac{1}{2}(\alpha_{i}-\alpha_{j})\right)}g\left(2r\sin{\left(\frac{1}{2}|\alpha_{i}-\alpha_{j}|\right)}\right)\right) \\
    &=2m_{i}m_{j}r\delta_{ij}\sin{\left(\frac{1}{2}(\alpha_{i}-\alpha_{j})\right)}g\left(2r\sin{\left(\frac{1}{2}|\alpha_{i}-\alpha_{j}|\right)}\right)\\
    &-2m_{i}m_{j}r^{2}\cos^{2}{\left(\frac{1}{2}(\alpha_{i}-\alpha_{j})\right)}g'\left(2r\sin{\left(\frac{1}{2}|\alpha_{i}-\alpha_{j}|\right)}\right)
  \end{align*}
  and thus
  \begin{align}\label{alpha i alpha j prep}
    \frac{\partial^{2}V}{\partial\alpha_{i}^{2}}&=-2\sum\limits_{j=1,\textrm{ }j\neq i}^{n}m_{i}m_{j}r\delta_{ij}\sin{\left(\frac{1}{2}(\alpha_{i}-\alpha_{j})\right)}g\left(2r\sin{\left(\frac{1}{2}|\alpha_{i}-\alpha_{j}|\right)}\right)\nonumber\\
    &+2\sum\limits_{j=1,\textrm{ }j\neq i}^{n}m_{i}m_{j}r^{2}\cos^{2}{\left(\frac{1}{2}(\alpha_{i}-\alpha_{j})\right)}g'\left(2r\sin{\left(\frac{1}{2}|\alpha_{i}-\alpha_{j}|\right)}\right)\nonumber\\
    &=-\sum\limits_{j=1,\textrm{ }j\neq i}^{n}\frac{\partial^{2}V}{\partial\alpha_{i}\partial\alpha_{j}},
  \end{align}
  so by (\ref{alpha i alpha j prep}),
  \begin{align*}
    &\sum\limits_{i=1}^{n}\sum\limits_{j=1}^{n}\gamma_{i}\gamma_{j}\frac{\partial^{2}V}{\partial\alpha_{i}\partial\alpha_{j}}=\sum\limits_{i=1}^{n}\gamma_{i}^{2}\frac{\partial^{2}V}{\partial\alpha_{i}^{2}}+\sum\limits_{i=1}^{n}\sum\limits_{j=1,\textrm{ }j\neq i}^{n}\gamma_{i}\gamma_{j}\frac{\partial^{2}V}{\partial\alpha_{i}\partial\alpha_{j}}\nonumber\\
    &=-\sum\limits_{i=1}^{n}\sum\limits_{j=1,\textrm{ }j\neq i}^{n}\gamma_{i}^{2}\frac{\partial^{2}V}{\partial\alpha_{i}\partial\alpha_{j}}+\sum\limits_{i=1}^{n}\sum\limits_{j=1,\textrm{ }j\neq i}^{n}\gamma_{i}\gamma_{j}\frac{\partial^{2}V}{\partial\alpha_{i}\partial\alpha_{j}}=\sum\limits_{i=1}^{n}\sum\limits_{j=1,\textrm{ }j\neq i}^{n}(-\gamma_{i}^{2}+\gamma_{i}\gamma_{j})\frac{\partial^{2}V}{\partial\alpha_{i}\partial\alpha_{j}},
  \end{align*}
  giving
  \begin{align*}
    &\sum\limits_{i=1}^{n}\sum\limits_{j=1}^{n}\gamma_{i}\gamma_{j}\frac{\partial^{2}V}{\partial\alpha_{i}\partial\alpha_{j}}=\sum\limits_{i=1}^{n}\sum\limits_{j=1,\textrm{ }j\neq i}^{n}(-\gamma_{i}^{2}+\gamma_{i}\gamma_{j})\frac{\partial^{2}V}{\partial\alpha_{i}\partial\alpha_{j}}\\
    &=\frac{1}{2}\left(\sum\limits_{i=1}^{n}\sum\limits_{j=1,\textrm{ }j\neq i}^{n}(-\gamma_{i}^{2}+\gamma_{i}\gamma_{j})\frac{\partial^{2}V}{\partial\alpha_{i}\partial\alpha_{j}}+\sum\limits_{j=1}^{n}\sum\limits_{i=1,\textrm{ }i\neq j}^{n}(-\gamma_{j}^{2}+\gamma_{j}\gamma_{i})\frac{\partial^{2}V}{\partial\alpha_{j}\partial\alpha_{i}}\right)\\
    &=-\frac{1}{2}\sum\limits_{i=1}^{n}\sum\limits_{j=1,\textrm{ }j\neq i}^{n}\left(\gamma_{i}^{2}-2\gamma_{i}\gamma_{j}+\gamma_{j}^{2}\right)\frac{\partial^{2}V}{\partial\alpha_{i}\partial\alpha_{j}}=-\frac{1}{2}\sum\limits_{i=1}^{n}\sum\limits_{j=1,\textrm{ }j\neq i}^{n}\left(\gamma_{i}-\gamma_{j}\right)^{2}\frac{\partial^{2}V}{\partial\alpha_{i}\partial\alpha_{j}}\end{align*}
    and therefore
    \begin{align}\label{Third part of THE INEQUALITY}
    \sum\limits_{i=1}^{n}\sum\limits_{j=1}^{n}\gamma_{i}\gamma_{j}\frac{\partial^{2}V}{\partial\alpha_{i}\partial\alpha_{j}}=-\sum\limits_{i=1}^{n}\sum\limits_{j=1,\textrm{ }j\neq i}^{n}(\gamma_{i}-\gamma_{j})^{2}m_{i}m_{j}r\sin{\left(\frac{1}{2}|\alpha_{i}-\alpha_{j}|\right)}g\left(2r\sin{\left(\frac{1}{2}|\alpha_{i}-\alpha_{j}|\right)}\right)\nonumber\\
    +\sum\limits_{j=1,\textrm{ }j\neq i}^{n}m_{i}m_{j}(\gamma_{i}-\gamma_{j})^{2}r^{2}\cos^{2}{\left(\frac{1}{2}(\alpha_{i}-\alpha_{j})\right)}g'\left(2r\sin{\left(\frac{1}{2}|\alpha_{i}-\alpha_{j}|\right)}\right).
  \end{align}
  Combining (\ref{THE INEQUALITY}), (\ref{First and second part of THE INEQUALITY}) and (\ref{Third part of THE INEQUALITY}), we now get that
  \begin{align*}
    &\rho^{2}\frac{\partial^{2}V}{\partial r^{2}}+2\rho\sum\limits_{i=1}^{n}\gamma_{i}\frac{\partial^{2}V}{\partial r\partial\alpha_{i}}+\sum\limits_{i=1}^{n}\sum\limits_{j=1}^{n}\gamma_{i}\gamma_{j}\frac{\partial^{2}V}{\partial\alpha_{i}\partial\alpha_{j}}=-\rho^{2}\sum\limits_{i=1}^{n}m_{i}A^{2}\\
    &+\sum\limits_{i=1}^{n}\sum\limits_{j=1,\textrm{ }j\neq i}^{n}m_{i}m_{j}\left(2\rho\sin{\left(\frac{1}{2}(\alpha_{i}-\alpha_{j})\right)}+r(\gamma_{i}-\gamma_{j})\cos{\left(\frac{1}{2}(\alpha_{i}-\alpha_{j})\right)}\right)^{2}\\
    &\cdot g'\left(2r\sin{\left(\frac{1}{2}|\alpha_{i}-\alpha_{j}|\right)}\right)+0\\
    &-\sum\limits_{i=1}^{n}\sum\limits_{j=1,\textrm{ }j\neq i}^{n}(\gamma_{i}-\gamma_{j})^{2}m_{i}m_{j}r\sin{\left(\frac{1}{2}|\alpha_{i}-\alpha_{j}|\right)}g\left(2r\sin{\left(\frac{1}{2}|\alpha_{i}-\alpha_{j}|\right)}\right).
  \end{align*}
  As by construction $g'\left(2r\sin{\left(\frac{1}{2}|\alpha_{i}-\alpha_{j}|\right)}\right)<0$ and $g\left(2r\sin{\left(\frac{1}{2}|\alpha_{i}-\alpha_{j}|\right)}\right)>0$, this means that
  \begin{align}
    &\rho^{2}\frac{\partial^{2}V}{\partial r^{2}}+\rho\sum\limits_{i=1}^{n}\gamma_{i}\frac{\partial^{2}V}{\partial r\partial\alpha_{i}}+\sum\limits_{i=1}^{n}\sum\limits_{j=1}^{n}\gamma_{i}\gamma_{j}\frac{\partial^{2}V}{\partial\alpha_{i}\partial\alpha_{j}}\leq 0
  \end{align}
  with equality if and only if $\gamma_{i}=\gamma_{j}$ and $\rho=0$, which can be prevented by fixing one of the $Q_{i}$, $i\in\{1,...,n\}$.
  This proves that any stationary point of $V$ gives a maximum value of $V$, proving Theorem~\ref{Main Theorem}.
  \section{Proof of Theorem~\ref{Main Theorem 2}}\label{Proof of main theorem 2}
    Let $p_{1}$,...,$p_{n}$ be a polygonal relative equilibrium of (\ref{EquationsOfMotion Curved}) and let for $i\in\{1,...,n\}$
     \begin{align}\label{Expression Pi}
       P_{i}=\rho\begin{pmatrix}
         \cos{\gamma_{i}}\\
         \sin{\gamma_{i}}
       \end{pmatrix},
     \end{align}
     where $\gamma_{1},...,\gamma_{n}\in[0,2\pi)$ are ordered from smallest to largest and $\rho>0$.
     We will prove that the $P_{i}$ will give rise to a system of equations in the same way the $Q_{i}$ in the proof of Theorem~\ref{Main Theorem} give rise to (\ref{Identity relative equilibria}):
     Inserting (\ref{Expression Pi}) into (\ref{EquationsOfMotion Curved}) and multiplying both sides of the resulting equation for the first two entries of $\ddot{p}_{i}$ with $T(-Bt)$ gives
     \begin{align*}
       -B^{2}P_{i}=\sum\limits_{j=1,\textrm{ }j\neq i}^{n}\frac{\widehat{m}_{j}(P_{j}-\sigma(\langle P_{i},P_{j}\rangle+\sigma z^{2})P_{i})}{(\sigma-\sigma(\langle P_{i},P_{j}\rangle+\sigma z^{2})^{2})^{\frac{3}{2}}}-\sigma(\dot{p}_{i}\odot\dot{p}_{i})P_{i},\textrm{ }i\in\{1,...,n\},
     \end{align*}
     which can be rewritten as
     \begin{align}\label{AlmostThereCurved}
       -B^{2}P_{i}&=\sum\limits_{j=1,\textrm{ }j\neq i}^{n}\frac{\widehat{m}_{j}(P_{j}-P_{i})}{(\sigma-\sigma(\langle P_{i},P_{j}\rangle+\sigma z^{2})^{2})^{\frac{3}{2}}}\nonumber\\
       &+\left(\sum\limits_{j=1,\textrm{ }j\neq i}^{n}\frac{\widehat{m}_{j}(1-\sigma(\langle P_{i},P_{j}\rangle+\sigma z^{2}))}{(\sigma-\sigma(\langle P_{i},P_{j}\rangle+\sigma z^{2})^{2})^{\frac{3}{2}}}-\sigma(\dot{p}_{i}\odot\dot{p}_{i})\right)P_{i}
     \end{align}
     and the identity for the third entry of $\ddot{p}_{i}$ then is
     \begin{align}\label{AlmostThereCurved z}
       0=\sum\limits_{j=1,\textrm{ }j\neq i}^{n}\frac{\widehat{m}_{j}(1-\sigma(\langle P_{i},P_{j}\rangle+\sigma z^{2}))z}{(\sigma-\sigma(\langle P_{i},P_{j}\rangle+\sigma z^{2})^{2})^{\frac{3}{2}}}-\sigma(\dot{p}_{i}\odot\dot{p}_{i})z.
     \end{align}
     Note that $-1=p_{i}\odot p_{i}=\|P_{i}\|^{2}-z^{2}=\rho^{2}-z^{2}$ for $\sigma=-1$, so $z\neq 0$. Therefore, using (\ref{AlmostThereCurved z}), the second sum of (\ref{AlmostThereCurved}) may be replaced with zero, giving
     \begin{align}\label{AlmostThereCurved2}
       -B^{2}P_{i}=\sum\limits_{j=1,\textrm{ }j\neq i}^{n}\frac{\widehat{m}_{j}(P_{j}-P_{i})}{(\sigma -\sigma(\langle P_{i},P_{j}\rangle+\sigma z^{2})^{2})^{\frac{3}{2}}},
     \end{align}
     which by (\ref{Expression Pi}) can be rewritten, using that $\sigma z^{2}=\sigma-\rho^{2}$ and multiplying both sides of (\ref{AlmostThereCurved2}) with $-T(-\gamma_{i})$, as
     \begin{align}\label{AlmostThereCurved3}
       B^{2}\begin{pmatrix}
         \rho\\ 0
       \end{pmatrix}&=\sum\limits_{j=1,\textrm{ }j\neq i}^{n}\widehat{m}_{j}\rho\begin{pmatrix}
         1-\cos{(\gamma_{j}-\gamma_{j})}\\ \sin{(\gamma_{j}-\gamma_{i})}
       \end{pmatrix}\nonumber\\
       &\cdot\rho^{-3}(1-\cos{(\gamma_{j}-\gamma_{i})})^{-\frac{3}{2}}(2-\sigma\rho^{2}(1-\cos{(\gamma_{j}-\gamma_{i})}))^{-\frac{3}{2}}.
     \end{align}
     As $\rho(1-\cos{(\gamma_{j}-\gamma_{i})})^{\frac{1}{2}}=\sqrt{2}\cdot \rho\sin\left(\frac{1}{2}|\gamma_{j}-\gamma_{i}|\right)$, we may rewrite (\ref{AlmostThereCurved3}) as
     \begin{align}\label{AlmostThereCurved4}
       B^{2}\begin{pmatrix}
         \rho\\ 0
       \end{pmatrix}&=\sum\limits_{j=1,\textrm{ }j\neq i}^{n}\widehat{m}_{j}\rho\begin{pmatrix}
         1-\cos{(\gamma_{j}-\gamma_{j})}\\ \sin{(\gamma_{j}-\gamma_{i})}
       \end{pmatrix}\nonumber\\
       &\cdot8\left(2\rho\sin\left(\frac{1}{2}|\gamma_{j}-\gamma_{i}|\right)\right)^{-3}\left(4-\sigma\left(2\rho\sin\left(\frac{1}{2}|\gamma_{j}-\gamma_{i}|\right)\right)^{2}\right)^{-\frac{3}{2}}.
     \end{align}
     If we define $h(x)=8x^{-3}(4-\sigma x^{2})^{-\frac{3}{2}}$, then we can rewrite (\ref{AlmostThereCurved4}) as
     \begin{align}\label{AlmostThereCurved5}
       B^{2}\begin{pmatrix}
         \rho\\ 0
       \end{pmatrix}&=\sum\limits_{j=1,\textrm{ }j\neq i}^{n}\widehat{m}_{j}\rho\begin{pmatrix}
         1-\cos{(\gamma_{j}-\gamma_{j})}\\ \sin{(\gamma_{j}-\gamma_{i})}
       \end{pmatrix}h\left(2\rho\sin\left(\frac{1}{2}|\gamma_{j}-\gamma_{i}|\right)\right).
     \end{align}
     Now (\ref{AlmostThereCurved5}) is an identity exactly of the same type as (\ref{Identity relative equilibria}), as $(xh(x))'<0$. So going through the proof of Theorem~\ref{Main Theorem} using (\ref{AlmostThereCurved5}) instead of (\ref{Identity relative equilibria}) now proves our theorem. It should be noted that (\ref{AlmostThereCurved3}) was already proven in \cite{D2} in a more general setting, but as the calculation is not that long for this specific case, the argument has been repeated to make the paper self contained.
    \section{Proof of Theorem~\ref{Main Theorem 3}}\label{Proof of main theorem 3}
     Let $n=N+1$. If $q_{1}$,...,$q_{n}$ is a relative equilibrium of (\ref{Equations of motion}) with $Q_{M}=0$, the point masses $q_{1}$,...,$q_{N}$ lie on a circle and $Q_{N+1}=0$, then we may write for $i\in\{1,...,N\}$
     \begin{align*}
       Q_{i}=r\begin{pmatrix}
         \cos{\alpha_{i}} \\ \sin{\alpha_{i}},
       \end{pmatrix}
     \end{align*}
     where $r>0$ and $0\leq\alpha_{1}<...<\alpha_{N}<2\pi$. Following the proof of Theorem~\ref{Main Theorem}, again for $i\in\{1,...,N\}$, inserting these expressions for the $Q_{i}$ into (\ref{Equations of motion}) gives instead of (\ref{Identity relative equilibria}) the slightly different identity
     \begin{align*}
       -A^{2}Q_{i}=\sum\limits_{j=1,\textrm{ }j\neq i}^{N}m_{j}(Q_{j}-Q_{i})f(\|Q_{j}-Q_{i}\|)+m_{N+1}(0-Q_{i})f(\|0-Q_{i}\|),
     \end{align*}
     giving
     \begin{align*}
       \left(m_{n}f(\|Q_{i}\|)-A^{2}\right)Q_{i}=\sum\limits_{j=1,\textrm{ }j\neq i}^{N}m_{j}(Q_{j}-Q_{i})f(\|Q_{j}-Q_{i}\|),
     \end{align*}
     which by the same argument that gave (\ref{Identity relative equilibria3}) now gives
     \begin{align}\label{Relative equilibrium with central mass}
       \begin{cases}
         m_{i}\left(A^{2}r-m_{n}g(r)\right)=\sum\limits_{j=1,\textrm{ }j\neq i}^{N}m_{i}m_{j}
      \sin{\left(\frac{1}{2}|\alpha_{i}-\alpha_{j}|\right)}
    g\left(2r\sin{\left(\frac{1}{2}|\alpha_{i}-\alpha_{j}|\right)}\right)\\
    0=\sum\limits_{j=1,\textrm{ }j\neq i}^{N}m_{i}m_{j}r\delta_{ij}\cos{\left(\frac{1}{2}(\alpha_{i}-\alpha_{j})\right)}g\left(2r\sin{\left(\frac{1}{2}|\alpha_{i}-\alpha_{j}|\right)}\right),
    \end{cases}
     \end{align}
     with again $g(x)=xf(x)$ and $\delta_{ij}$ as in the proof of Theorem~\ref{Main Theorem}. So as long as $\left(A^{2}-m_{n}f(r)\right)>0$ (as the right hand side of the first identity of (\ref{Relative equilibrium with central mass}) has to be positive) we can continue to go through the proof of Theorem~\ref{Main Theorem}, replacing the function $V$ (see (\ref{definition V})) with
     \begin{align*}W(r,\alpha_{1},...,\alpha_{N})=\sum\limits_{l=1}^{N}\sum\limits_{k=1,\textrm{ }k\neq l}^{N}m_{l}m_{k}G\left(2r\sin{\left(\frac{1}{2}|\alpha_{l}-\alpha_{k}|\right)}\right)-\sum\limits_{l=1}^{n}m_{l}\left(A^{2}r^{2}-2m_{n}G(r)\right).
     \end{align*}
     Repeating the proof of Theorem~\ref{Main Theorem} using $W$ instead of $V$ leads to
     \begin{align*}
       &\rho^{2}\frac{\partial^{2}W}{\partial r^{2}}+2\rho\sum\limits_{i=1}^{N}\gamma_{i}\frac{\partial^{2}W}{\partial r\partial\alpha_{i}}+\sum\limits_{i=1}^{N}\sum\limits_{j=1}^{N}\gamma_{i}\gamma_{j}\frac{\partial^{2}W}{\partial\alpha_{i}\partial\alpha_{j}}=-2\rho^{2}\sum\limits_{i=1}^{N}m_{i}(A^{2}-m_{n}g'(r))\\
       &+\sum\limits_{i=1}^{N}\sum\limits_{j=1,\textrm{ }j\neq i}^{N}m_{i}m_{j}\left(2\rho\sin{\left(\frac{1}{2}(\alpha_{i}-\alpha_{j})\right)}+r(\gamma_{i}-\gamma_{j})\cos{\left(\frac{1}{2}(\alpha_{i}-\alpha_{j})\right)}\right)^{2}\\
       &\cdot g'\left(2r\sin{\left(\frac{1}{2}|\alpha_{i}-\alpha_{j}|\right)}\right)\\
       &-\sum\limits_{i=1}^{N}\sum\limits_{j=1,\textrm{ }j\neq i}^{N}(\gamma_{i}-\gamma_{j})^{2}m_{i}m_{j}r\sin{\left(\frac{1}{2}|\alpha_{i}-\alpha_{j}|\right)}g\left(2r\sin{\left(\frac{1}{2}|\alpha_{i}-\alpha_{j}|\right)}\right).
    \end{align*}
    for all $\rho$, $\gamma_{1}$,...,$\gamma_{N}\in\mathbb{R}$.
    As $g'\left(2r\sin{\left(\frac{1}{2}|\alpha_{i}-\alpha_{j}|\right)}\right)<0$, $g'(r)<0$ and $g\left(2r\sin{\left(\frac{1}{2}|\alpha_{i}-\alpha_{j}|\right)}\right)>0$, this means that
     \begin{align*}
       &\rho^{2}\frac{\partial^{2}W}{\partial r^{2}}+2\rho\sum\limits_{i=1}^{N}\gamma_{i}\frac{\partial^{2}W}{\partial r\partial\alpha_{i}}+\sum\limits_{i=1}^{N}\sum\limits_{j=1}^{N}\gamma_{i}\gamma_{j}\frac{\partial^{2}W}{\partial\alpha_{i}\partial\alpha_{j}}\leq 0
     \end{align*}
     with equality if and only if $\rho=0$ and $\gamma_{i}=\gamma_{j}$ for all $i$, $j\in\{1,...,N\}$, which can be prevented by fixing one of the $Q_{i}$, $i\in\{1,...,N\}$. We thus find that by the same argument as used in the proof of Theorem~\ref{Main Theorem}, that for any order of masses $m_{1}$,...,$m_{n}$, there exists at most one relative equilibrium of (\ref{Equations of motion}) with center of mass zero, where all point masses but one lie on a circle around the remaining point mass, which coincides with the center of mass.

\end{document}